\def\rfr#1{eq. (\ref{#1})}

\def\bar{\begin{eqnarray}}
\def\ear{\end{eqnarray}}
\def\bb{\bibitem}
\def\eqi{\begin{equation}}
\def\eqf{\end{equation}}
\def\eqia{\begin{eqnarray}}
\def\eqfa{\end{eqnarray}}
\def\rp#1#2{{#1\over#2}}

\def\lb#1{\label{#1}}

\def\virg#1{``#1''}

\def\oc2{$\mathcal{O}(c^{-2})$}

\documentclass[CEJP,PDF]{cej} % use PDF command to enable PDFLaTeX driver
\usepackage{layout}
\usepackage{amsmath}
\usepackage{textcomp}

\title{Conservative evaluation of the uncertainty in the
    LAGEOS-LAGEOS II Lense-Thirring test}

\articletype{Research Article}

\author{
Lorenzo~Iorio\inst{1}\email{lorenzo.iorio@libero.it}
}

\institute{
     \inst{1} INFN-Sezione di Pisa,\\
     Viale Unit$\grave{\rm a}$ di Italia 68, 70125 Bari, Italy
          }

\abstract{
  We deal with the test of the general relativistic gravitomagnetic
  Lense-Thirring effect currently being conducted in the Earth's
  gravitational field with the combined nodes $\Omega$ of the
  laser-ranged geodetic satellites LAGEOS and LAGEOS II.
  One of the most important sources of systematic uncertainty on the
  orbits of the LAGEOS satellites, with respect to the Lense-Thirring
  signature, is the bias due to the even zonal harmonic coefficients
  $J_{\ell}$ of the multipolar expansion of the Earth's geopotential
  which account for the departures from sphericity of the terrestrial
  gravitational potential induced by the centrifugal effects of its
  diurnal rotation.  The issue addressed here is: are the so far
  published evaluations of such a systematic error reliable and
  realistic?  The answer is negative.  Indeed, if the difference
  $\Delta J_{\ell}$ among the even zonals estimated in different
  global solutions (EIGEN-GRACE02S, EIGEN-CG03C, GGM02S, GGM03S,
  ITG-Grace02, ITG-Grace03s, JEM01-RL03B, EGM2008, AIUB-GRACE01S) is
  assumed for the uncertainties $\delta J_{\ell}$ instead of using
  their more-or-less calibrated covariances $\sigma_{J_\ell}$, it
  turns out that the systematic error $\delta\mu$ in the
  Lense-Thirring measurement is about 3 to 4 times larger than in the
  evaluations so far published based on the use of the covariances of
  one model at a time separately, amounting up to $37\%$ for the pair
  EIGEN-GRACE02S/ITG-Grace03s. The comparison among the other recent
  GRACE-based models yields bias as large as about $25-30\%$.  The
  major discrepancies still occur for $J_4, J_6$ and $J_8$, which are
  just to which the zonals the combined LAGEOS/LAGOES II nodes are
  most sensitive.
                                   }


\begin{document}
\maketitle

%% ###################################################################
%
\section{Introduction}
In the weak-field and slow motion approximation, the Einstein field
equations of general relativity get linearized to a form resembling
Maxwell's equations of electromagnetism. Thus, a gravitomagnetic
field, induced by the off-diagonal components $g_{0i}, i=1,2,3$ of the
space-time metric tensor related to the mass-energy currents of the
source of the gravitational field, arises \cite{MashNOVA}. It affects
in several ways the motion of, e.g., test particles and
electromagnetic waves \cite{Rug}. Perhaps the most famous
gravitomagnetic effects are gyroscope precession \cite{Pugh,Schi} and
the Lense-Thirring\footnote{According to an interesting historical
  analysis recently performed in Ref. \cite{Pfi07}, it would be more
  correct to speak about an Einstein-Thirring-Lense effect.}
precessions \cite{LT} of the orbit of a test particle, both occurring
in the field of a central slowly rotating mass like a planet.

The measurement of gyroscope precession in the Earth's gravitational
field has been the goal of the dedicated space-based GP-B
mission\footnote{See \url{http://einstein.stanford.edu/}} \cite{Eve,
  GPB} launched in 2004; its data analysis is still in progress.

In this paper we critically discuss some issues concerning the test of
the Lense-Thirring effect performed with the LAGEOS and LAGEOS II
terrestrial artificial satellites \cite{Ciu04} tracked with the
Satellite Laser Ranging (SLR) technique \cite{Pearl02}.

\cite{vpe76a,vpe76b} proposed measuring the Lense-Thirring nodal
precession of a pair of counter-orbiting spacecraft in terrestrial
polar orbits and equipped with drag-free apparatus.  A somewhat
equivalent, cheaper version of such an idea was put forth in Ref.
\cite{Ciu86} whose author proposed to launch a passive, geodetic satellite in
an orbit identical to that of the LAGEOS satellite apart from the
orbital planes which should have been displaced by 180 deg
apart.\footnote{ LAGEOS was put into orbit in 1976, followed by its
  twin LAGEOS II in 1992.}  The measurable quantity was, in this case,
the sum of the nodes of LAGEOS and of the new spacecraft, later named
LAGEOS III, LARES, WEBER-SAT, in order to cancel the confounding
effects of the multipoles of the Newtonian part of the terrestrial
gravitational potential (see below). Although extensively studied by
various groups \cite{CSR,LARES,Ioretal02}, such an idea has not been
implemented for a long time.  Recently, the Italian Space Agency (ASI)
has approved this project and should launch a VEGA rocket for this
purpose at the end of 2009-beginning of 2010
(\url{http://www.asi.it/en/activity/cosmology/lares}).  For recent
updates of the LARES/WEBER-SAT mission, including recently added
additional goals in fundamental physics and related criticisms, see  Refs.
\cite{LucPao01,Ior02,Ciu04b,Ciu06b,IorJCAP,Ior07c,Ior07d,Ior09}.

Among scenarios involving \emph{existing} orbiting bodies, the idea of
measuring the Lense-Thirring node rate with the just launched LAGEOS
satellite, along with the other SLR targets orbiting at that time, was
proposed in Ref. \cite{Cug78}. Tests have been effectively performed using
the LAGEOS and LAGEOS II satellites \cite{tanti}, according to a
strategy \cite{Ciu96} involving a suitable combination of the nodes
of both satellites and the perigee $\omega$ of LAGEOS II. This was
done to reduce the impact of the most relevant source of systematic
bias, viz. the mismodelling in the even ($\ell=2,4,6\ldots$) zonal
($m=0$) harmonics $J_{\ell}$ of the multipolar expansion of the
Newtonian part of the terrestrial gravitational
potential:\footnote{The relation among the even zonals $J_{\ell}$ and
  the normalized gravity coefficients $\overline{C}_{\ell 0}$ is
  $J_{\ell}=-\sqrt{2\ell + 1}\ \overline{C}_{\ell 0}$.} they account
for non-sphericity of the terrestrial gravitational field induced by
centrifugal effects of the Earth's diurnal rotation. The even zonals
affect the node and the perigee of a terrestrial satellite with
secular precessions which may mimic the Lense-Thirring signature. The
three-elements combination used allowed for removing the uncertainties
in $J_2$ and $J_4$. In \cite{Ciu98} a $\approx 20\%$ test was
reported by using the\footnote{Contrary to the subsequent
  CHAMP/GRACE-based models, EGM96 relies upon multidecadal tracking of
  SLR data of a constellation of geodetic satellites including LAGEOS
  and LAGEOS II as well; thus the possibility of a sort of $a-priori$
  `imprinting' of the Lense-Thirring effect itself, not solved-for in
  EGM96, cannot be neglected.} EGM96 \cite{Lem98} Earth gravity
model; subsequent analyses showed that such an evaluation of the total
error budget was overly optimistic in view of the likely unreliable
computation of the total bias due to the even zonals
\cite{Ior03,Ries03a,Ries03b}.  An analogous, huge underestimation
turned out to hold also for the effect of non-gravitational
perturbations \cite{Mil87} like direct solar radiation pressure, the
Earth's albedo, various subtle thermal effects depending on the the
physical properties of the satellites' surfaces and their rotational
state \cite{Inv94,Ves99,Luc01,Luc02,Luc03,Luc04,Lucetal04,Ries03a},
which the perigees of LAGEOS-like satellites are particularly
sensitive to. As a result, the realistic total error budget in the
test reported in Ref. \cite{Ciu98} might be as large as $60-90\%$ or even
more (by considering EGM96 only).

The observable used in Ref. \cite{Ciu04} with the GRACE-only
EIGEN-GRACE02S model \cite{eigengrace02s} and in Ref. \cite{Ries08} with
other global terrestrial gravity solutions was the following linear
combination\footnote{See also \cite{Pav02,Ries03a,Ries03b}.} of the
nodes of LAGEOS and LAGEOS II, explicitly computed in Ref. \cite{IorMor}
following the approach proposed in Ref. \cite{Ciu96}:
\begin{equation} f=\dot\Omega^{\rm LAGEOS}+ c_1\dot\Omega^{\rm LAGEOS\
    II }, \lb{combi}\end{equation} where \begin{equation}
  c_1\equiv-\rp{\dot\Omega^{\rm LAGEOS}_{.2}}{\dot\Omega^{\rm LAGEOS\
      II }_{.2}}=-\rp{\cos i_{\rm LAGEOS}}{\cos i_{\rm LAGEOS\
      II}}\left(\rp{1-e^2_{\rm LAGEOS\ II}}{1-e^2_{\rm
        LAGEOS}}\right)^2\left(\rp{a_{\rm LAGEOS\ II}}{a_{\rm
        LAGEOS}}\right)^{7/2}.\lb{coff}\end{equation} The coefficients
$\dot\Omega_{.\ell}$ of the aliasing classical node precessions
\cite{Kau} $\dot\Omega_{\rm
  class}=\sum_{\ell}\dot\Omega_{.\ell}J_{\ell}$ induced by even zonals
have been analytically worked out in e.g. \cite{Ior03}; $a,e,i$ are
the satellite's semimajor axis, eccentricity and inclination,
respectively and yield $c_1=0.544$ for \rfr{coff}.  The Lense-Thirring
signature of \rfr{combi} amounts to 47.8 milliarcseconds per year (mas
yr$^{-1}$). The combination \rfr{combi} allows, by construction, to
remove the aliasing effects due to the static and time-varying parts
of the first even zonal $J_2$. The nominal bias (computed with the
estimated values of $J_{\ell}$, $\ell=4,6...$) due to the remaining
higher degree even zonals would amount to about $10^5$ mas~yr$^{-1}$;
the need of a careful and reliable modeling of such an important
source of systematic bias is, thus, quite apparent. Conversely, the
nodes of the LAGEOS-type spacecraft are affected by the
non-gravitational accelerations $\approx 1\%$ of the
Lense-Thirring effect \cite{Luc01,Luc02,Luc03,Luc04,Lucetal04}. For a
comprehensive, up-to-date overview of the numerous and subtle issues
concerning the measurement of the Lense-Thirring effect see
\cite{IorNOVA}.

Here, we will address the following questions:
\begin{itemize}
\item Has the systematic error due to the competing secular node
  precessions induced by the static part of the even zonal harmonics
  been realistically evaluated so far in literature? (Section
  \ref{grav})
  % \item Why has the analysis with the LAGEOS satellites not been
  %   repeated so far by any other independent team? (\sref{repeat})
\item Are other approaches to extract the gravitomagnetic signature
  from the data feasible? (Section \ref{approach})
\end{itemize}

\section{The systematic error of gravitational origin}\lb{grav}
The realistic evaluation of the total error budget of such a test
raised a lively debate
% \footnote{It may be interesting to look also at the history and
%   discussion sections of the item frame dragging at
%   http://en.wikipedia.org
%   with %particular care at the systematic censorship of certain specific bibliographic references occurred in January-April 2007.}
\cite{Ciu05,Ciu06,IorNA,IorJoG,IorGRG,Ior07,Luc05}, mainly focussed
on the impact of the static and time-varying parts of the Newtonian
component of the Earth's gravitational potential through the aliasing
secular precessions induced on a satellite's node.  A common feature
of all the competing evaluations so far published is that the
systematic bias due to the static component of the geopotential was
calculated always by using the released (more or less accurately
calibrated) covariances $\sigma_{J_{\ell}}$ of one Earth gravity model
solution at a time for the uncertainties $\delta J_{\ell}$ in the even
zonal harmonics, yielding a percentage error particular to each model.

Since it is always difficult to reliably calibrate the formal, statistical uncertainties
in the estimated zonals of the covariance matrix for a global solution, it is much more realistic and conservative to instead take
the differences\footnote{See Fig. 5 of \cite{Luc07} for a comparison of
  the estimated $\overline{C}_{40}$ in different models.} $\Delta
J_{\ell}$ between the estimated even zonals for different pairs of Earth
gravity field solutions as representative of the real uncertainty
$\delta J_{\ell}$ in the zonals \cite{Lerch}. In Table
\ref{tavola1}--Table \ref{tavolaAIUB2} we present our results for the
most recent GRACE-based models released so far by different
institutions and retrievable on the Internet at\footnote{I thank J
  Ries, CSR, and M Watkins (JPL) for having provided me with the even
  zonals of the GGM03S \cite{ggm03} and JEM01-RL03B models.}
http://icgem.gfz-potsdam.de/ICGEM/ICGEM.html.  The models used are
EIGEN-GRACE02S \cite{eigengrace02s} and EIGEN-CG03C
\cite{eigencg03c} from GFZ (Potsdam, Germany), GGM02S \cite{ggm02}
and GGM03S \cite{ggm03} from CSR (Austin, Texas), ITG-Grace02s
\cite{ITG} and ITG-Grace03 \cite{itggrace03s} from IGG (Bonn,
Germany), JEM01-RL03B from JPL (NASA, USA), EGM2008 \cite{egm2008}
from NGA (USA) and AIUB-GRACE01S \cite{aiub} from AIUB (Bern,
Switzerland).  This approach was taken also in  Ref.
\cite{Ciu96} with the JGM3 and GEMT-2 models.  We included both the
sum of the absolute values (SAV) of each mismodelled term and the
square root of the sum of the squares (RSS) of each mismodelled term.
\begin{table}
  \caption{\label{tavola1}Impact of the mismodelling in the even zonal harmonics on $f_{\ell}=\left|\dot\Omega^{\rm LAGEOS}_{\ell} + c_1\dot\Omega^{\rm LAGEOS\ II}_{.\ell}\right|\Delta J_{\ell},\ \ell=4,\dots,20$, in mas yr$^{-1}$. Recall that $J_{\ell}=-\sqrt{2\ell + 1}\ \overline{C}_{\ell 0}$; for the uncertainty in the even zonals we have taken here the difference $\Delta\overline{C}_{\ell 0}=\left|\overline{C}_{\ell 0}^{\rm (X)}-\overline{C}_{\ell 0}^{\rm (Y)}\right|$ between the model X $=$ EIGEN-CG03C \cite{eigencg03c} and the model Y $=$ EIGEN-GRACE02S \cite{eigengrace02s}.
    EIGEN-CG03C combines data from CHAMP (860 days out of October 2000
    to June 2003), GRACE (376 days out of February to
    May 2003, July to December 2003 and February to July 2004) and terrestrial measurements; EIGEN-GRACE02S is based on 110 days (out of August and November 2002 and April, May and August 2003) of GRACE-only GPS-GRACE high-low satellite-to-satellite data, on-board measurements of non-gravitational accelerations, and especially GRACE intersatellite tracking data. $\sigma_{\rm X/Y}$ are the covariance calibrated errors for both models. Values of $f_{\ell}$ smaller than 0.1 mas yr$^{-1}$ have not been quoted. The Lense-Thirring precession of the combination of \rfr{combi} amounts to 47.8 mas yr$^{-1}$. The percent bias $\delta\mu$ has been computed by normalizing the linear sum of $f_{\ell}, \ell=4,\dots,20$ to the Lense-Thirring precession. The discrepancies between the models are significant since $\Delta \overline{C}_{\ell 0}$ are larger than the linearly added sigmas for $\ell=4,...16$.}
  \begin{tabular}{lccc}
    \hline
    $\ell$ & $\Delta\overline{C}_{\ell 0}$ (EIGEN-CG03C-EIGEN-GRACE02S) & $\sigma_{\rm  X}+\sigma_{\rm Y}$ & $f_{\ell}$  (mas yr$^{-1}$)\\
    \hline\hline
    % 2 & $4.81\times 10^{-11}$ & 0\\
    4 & $1.96\times 10^{-11}$ &  $1.01\times 10^{-11}$ & 7.3\\
    6 & $2.50\times 10^{-11}$ &  $4.8\times 10^{-12}$ & 5.4\\
    8 & $4.9\times 10^{-12}$ &  $3.3\times 10^{-12}$ & 0.2\\
    10 & $3.7\times 10^{-12}$ &  $3.4\times 10^{-12}$ & -\\
    12 & $2.5\times 10^{-12}$ &  $2.3\times 10^{-12}$ & -\\
    14 & $6.1\times 10^{-12}$ &  $2.1\times 10^{-12}$ & -\\
    16 & $2.1\times 10^{-12}$ &  $1.7\times 10^{-12}$ & -\\
    18 & $6\times 10^{-13}$ &  $1.7\times 10^{-12}$ & -\\
    20 & $1.7\times 10^{-12}$ &  $1.7\times 10^{-12}$ & -\\
    \hline
    &     $\delta\mu = 27\%$ (SAV) & $\delta\mu = 19\%$ (RSS) &   \\
    \hline
  \end{tabular}
\end{table}
\begin{table}
  \caption{Impact of the mismodelling in the even zonal harmonics as solved for in X=GGM02S \cite{ggm02} and  Y=ITG-Grace02s \cite{ITG}.
    GGM02S is based on 363 days of GRACE-only data   (GPS and intersatellite tracking, neither constraints nor regularization applied)
    spread between April 4, 2002 and Dec 31, 2003. The $\sigma$ are formal for both models. $\Delta \overline{C}_{\ell 0}$ are always larger than the linearly added sigmas, apart from   $\ell=12$ and $\ell=18$.}\label{tavola3}
  \begin{tabular}{lccc}
    \hline
    $\ell$ & $\Delta\overline{C}_{\ell 0}$ (GGM02S-ITG-Grace02s) & $\sigma_{\rm  X}+\sigma_{\rm Y}$ & $f_{\ell}$  (mas yr$^{-1}$)\\
    \hline\hline
    % 2 & $4.81\times 10^{-11}$ & 0\\
    4 & $1.9\times 10^{-11}$ &  $8.7\times 10^{-12}$ & 7.2\\
    6 & $2.1\times 10^{-11}$ &  $4.6\times 10^{-12}$ & 4.6\\
    8 & $5.7\times 10^{-12}$ &  $2.8\times 10^{-12}$ & 0.2\\
    10 & $4.5\times 10^{-12}$ &  $2.0\times 10^{-12}$ & -\\
    12 & $1.5\times 10^{-12}$ &  $1.8\times 10^{-12}$ & -\\
    14 & $6.6\times 10^{-12}$ &  $1.6\times 10^{-12}$ & -\\
    16 & $2.9\times 10^{-12}$ &  $1.6\times 10^{-12}$ & -\\
    18 & $1.4\times 10^{-12}$ &  $1.6\times 10^{-12}$ & -\\
    20 & $2.0\times 10^{-12}$ &  $1.6\times 10^{-12}$ & -\\

    \hline
    &    $\delta\mu = 25\%$ (SAV) & $\delta\mu = 18\%$ (RSS) &   \\  %
    \hline
  \end{tabular}
\end{table}
\begin{table}
  \caption{Impact of the mismodelling in the even zonal harmonics as solved for in X=GGM02S \cite{ggm02} and  Y=EIGEN-CG03C \cite{eigencg03c}.
    The $\sigma$ are formal for GGM02S, calibrated for EIGEN-CG03C. $\Delta \overline{C}_{\ell 0}$ are always larger than the linearly added sigmas.}\label{tavola3}
  \begin{tabular}{lccc}
    \hline
    $\ell$ & $\Delta\overline{C}_{\ell 0}$ (GGM02S-EIGEN-CG03C) & $\sigma_{\rm  X}+\sigma_{\rm Y}$ & $f_{\ell}$  (mas yr$^{-1}$)\\
    \hline\hline
    % 2 & $4.81\times 10^{-11}$ & 0\\
    4 & $1.81\times 10^{-11}$ &  $3.7\times 10^{-12}$ & 6.7\\
    6 & $1.53\times 10^{-11}$ &  $1.8\times 10^{-12}$ & 3.3\\
    8 & $1.5\times 10^{-12}$ &  $1.1\times 10^{-12}$ & -\\
    10 & $4.9\times 10^{-12}$ &  $8\times 10^{-13}$ & -\\
    12 & $8\times 10^{-13}$ &  $7\times 10^{-13}$ & -\\
    14 & $7.7\times 10^{-12}$ &  $6\times 10^{-13}$ & -\\
    16 & $3.8\times 10^{-12}$ &  $5\times 10^{-13}$ & -\\
    18 & $2.1\times 10^{-12}$ &  $5\times 10^{-13}$ & -\\
    20 & $2.3\times 10^{-12}$ &  $4\times 10^{-13}$ & -\\

    \hline
    &    $\delta\mu = 22\%$ (SAV) & $\delta\mu = 16\%$ (RSS) &   \\  %
    \hline
  \end{tabular}
\end{table}

\begin{table}
  \caption{Bias due to the mismodelling in the even zonals of the models X=ITG-Grace03s \cite{itggrace03s}, based on GRACE-only accumulated normal equations from data out of September 2002-April 2007 (neither apriori information nor regularization used), and Y=GGM02S \cite{ggm02}.  The $\sigma$ for both models are formal. $\Delta \overline{C}_{\ell 0}$ are always larger than the linearly added sigmas, apart from  $\ell=12$ and $\ell=18$.}\label{tavola11}
  \begin{tabular}{lccc}
    \hline
    $\ell$ & $\Delta\overline{C}_{\ell 0}$ (ITG-Grace03s-GGM02S) & $\sigma_{\rm  X}+\sigma_{\rm Y}$ & $f_{\ell}$  (mas yr$^{-1}$)\\
    \hline\hline
    % 2 & $4.81\times 10^{-11}$ & 0\\
    4 & $2.58\times 10^{-11}$ &  $8.6\times 10^{-12}$ & 9.6\\
    6 & $1.39\times 10^{-11}$ &  $4.7\times 10^{-12}$ & 3.1\\
    8 & $5.6\times 10^{-12}$ &  $2.9\times 10^{-12}$ & 0.2\\
    10 & $1.03\times 10^{-11}$ &  $2\times 10^{-12}$ & -\\
    12 & $7\times 10^{-13}$ &  $1.8\times 10^{-12}$ & -\\
    14 & $7.3\times 10^{-12}$ &  $1.6\times 10^{-12}$ & -\\
    16 & $2.6\times 10^{-12}$ &  $1.6\times 10^{-12}$ & -\\
    18 & $8\times 10^{-13}$ &  $1.6\times 10^{-12}$ & -\\
    20 & $2.4\times 10^{-12}$ &  $1.6\times 10^{-12}$ & -\\

    \hline
    &    $\delta\mu = 27\%$ (SAV) & $\delta\mu = 21\%$ (RSS) &   \\  %
    \hline
  \end{tabular}
\end{table}
\begin{table}
  \caption{Bias due to the mismodelling in the even zonals of the models  X = GGM02S \cite{ggm02} and Y = GGM03S \cite{ggm03} retrieved from data spanning January 2003 to December 2006.
    The $\sigma$ for GGM03S are calibrated. $\Delta \overline{C}_{\ell 0}$ are larger than the linearly added sigmas for $\ell = 4,6$. (The other zonals are of no concern)}\label{tavola03S}
  \begin{tabular}{lccc}
    \hline
    $\ell$ & $\Delta\overline{C}_{\ell 0}$ (GGM02S-GGM03S) & $\sigma_{\rm  X}+\sigma_{\rm Y}$ & $f_{\ell}$  (mas yr$^{-1}$)\\
    \hline\hline
    % 2 & $4.81\times 10^{-11}$ & 0\\
    4 & $1.87\times 10^{-11}$ &  $1.25\times 10^{-11}$ & 6.9\\
    6 & $1.96\times 10^{-11}$ &  $6.7\times 10^{-12}$ & 4.2\\
    8 & $3.8\times 10^{-12}$ &  $4.3\times 10^{-12}$ & 0.1\\
    10 & $8.9\times 10^{-12}$ &  $2.8\times 10^{-12}$ & 0.1\\
    12 & $6\times 10^{-13}$ &  $2.4\times 10^{-12}$ & -\\
    14 & $6.6\times 10^{-12}$ &  $2.1\times 10^{-12}$ & -\\
    16 & $2.1\times 10^{-12}$ &  $2.0\times 10^{-12}$ & -\\
    18 & $1.8\times 10^{-12}$ &  $2.0\times 10^{-12}$ & -\\
    20 & $2.2\times 10^{-12}$ &  $1.9\times 10^{-12}$ & -\\

    \hline
    &    $\delta\mu = 24\%$ (SAV) & $\delta\mu = 17\%$ (RSS) &   \\  %
    \hline
  \end{tabular}
\end{table}
\begin{table}
  \caption{Bias due to the mismodelling in the even zonals of the models  X = EIGEN-GRACE02S \cite{eigengrace02s} and Y = GGM03S \cite{ggm03}.
    The $\sigma$ for both models are calibrated. $\Delta \overline{C}_{\ell 0}$ are always larger than the linearly added sigmas apart from $\ell = 14,18$.}\label{tavola033S}
  \begin{tabular}{lccc}
    \hline
    $\ell$ & $\Delta\overline{C}_{\ell 0}$ (EIGEN-GRACE02S-GGM03S) & $\sigma_{\rm  X}+\sigma_{\rm Y}$ & $f_{\ell}$  (mas yr$^{-1}$)\\
    \hline\hline
    % 2 & $4.81\times 10^{-11}$ & 0\\
    4 & $2.00\times 10^{-11}$ &  $8.1\times 10^{-12}$ & 7.4\\
    6 & $2.92\times 10^{-11}$ &  $4.3\times 10^{-12}$ & 6.3\\
    8 & $1.05\times 10^{-11}$ &  $3.0\times 10^{-12}$ & 0.4\\
    10 & $7.8\times 10^{-12}$ &  $2.9\times 10^{-12}$ & 0.1\\
    12 & $3.9\times 10^{-12}$ &  $1.8\times 10^{-12}$ & -\\
    14 & $5\times 10^{-13}$ &  $1.7\times 10^{-12}$ & -\\
    16 & $1.7\times 10^{-12}$ &  $1.4\times 10^{-12}$ & -\\
    18 & $2\times 10^{-13}$ &  $1.4\times 10^{-12}$ & -\\
    20 & $2.5\times 10^{-12}$ &  $1.4\times 10^{-12}$ & -\\

    \hline
    &    $\delta\mu = 30\%$ (SAV) & $\delta\mu = 20\%$ (RSS) &   \\  %
    \hline
  \end{tabular}
\end{table}
\begin{table}
  \caption{Bias due to the mismodelling in the even zonals of the models  X = JEM01-RL03B, based on 49 months of GRACE-only data, and Y = GGM03S \cite{ggm03}.
    The $\sigma$ for GGM03S are calibrated. $\Delta \overline{C}_{\ell 0}$ are always larger than the linearly added sigmas apart from $\ell = 16$.}\label{tavolaJEM1}
  \begin{tabular}{lccc}
    \hline
    $\ell$ & $\Delta\overline{C}_{\ell 0}$ (JEM01-RL03B-GGM03S) & $\sigma_{\rm  X}+\sigma_{\rm Y}$ & $f_{\ell}$  (mas yr$^{-1}$)\\
    \hline\hline
    % 2 & $4.81\times 10^{-11}$ & 0\\
    4 & $1.97\times 10^{-11}$ &  $4.3\times 10^{-12}$ & 7.3\\
    6 & $2.7\times 10^{-12}$ &  $2.3\times 10^{-12}$ & 0.6\\
    8 & $1.7\times 10^{-12}$ &  $1.6\times 10^{-12}$ & -\\
    10 & $2.3\times 10^{-12}$ &  $8\times 10^{-13}$ & -\\
    12 & $7\times 10^{-13}$ &  $7\times 10^{-13}$ & -\\
    14 & $1.0\times 10^{-12}$ &  $6\times 10^{-13}$ & -\\
    16 & $2\times 10^{-13}$ &  $5\times 10^{-13}$ & -\\
    18 & $7\times 10^{-13}$ &  $5\times 10^{-13}$ & -\\
    20 & $5\times 10^{-13}$ &  $4\times 10^{-13}$ & -\\

    \hline
    &    $\delta\mu = 17\%$ (SAV) & $\delta\mu = 15\%$ (RSS) &   \\  %
    \hline
  \end{tabular}
\end{table}
\begin{table}
  \caption{Bias due to the mismodelling in the even zonals of the models  X = JEM01-RL03B and Y = ITG-Grace03s \cite{itggrace03s}.
    The $\sigma$ for ITG-Grace03s are formal. $\Delta \overline{C}_{\ell 0}$ are always larger than the linearly added sigmas.}\label{tavolaJEM2}
  \begin{tabular}{lccc}
    \hline
    $\ell$ & $\Delta\overline{C}_{\ell 0}$ (JEM01-RL03B-ITG-Grace03s) & $\sigma_{\rm  X}+\sigma_{\rm Y}$ & $f_{\ell}$  (mas yr$^{-1}$)\\
    \hline\hline
    % 2 & $4.81\times 10^{-11}$ & 0\\
    4 & $2.68\times 10^{-11}$ &  $4\times 10^{-13}$ & 9.9\\
    6 & $3.0\times 10^{-12}$ &  $2\times 10^{-13}$ & 0.6\\
    8 & $3.4\times 10^{-12}$ &  $1\times 10^{-13}$ & 0.1\\
    10 & $3.6\times 10^{-12}$ &  $1\times 10^{-13}$ & -\\
    12 & $6\times 10^{-13}$ &  $9\times 10^{-14}$ & -\\
    14 & $1.7\times 10^{-12}$ &  $9\times 10^{-14}$ & -\\
    16 & $4\times 10^{-13}$ &  $8\times 10^{-14}$ & -\\
    18 & $4\times 10^{-13}$ &  $8\times 10^{-14}$ & -\\
    20 & $7\times 10^{-13}$ &  $8\times 10^{-14}$ & -\\

    \hline
    &    $\delta\mu = 22\%$ (SAV) & $\delta\mu = 10\%$ (RSS) &   \\  %
    \hline
  \end{tabular}
\end{table}
\begin{table}
  \caption{Aliasing effect of the mismodelling in the even zonal harmonics estimated in the X=ITG-Grace03s \cite{itggrace03s} and the Y=EIGEN-GRACE02S \cite{eigengrace02s} models.  The covariance matrix $\sigma$ for ITG-Grace03s are formal, while the ones of EIGEN-GRACE02S are calibrated. $\Delta \overline{C}_{\ell 0}$ are larger than the linearly added sigmas for $\ell =4,...,20$, apart from $\ell=18$. }\label{tavola7}
  \begin{tabular}{lccc}
    \hline
    $\ell$ & $\Delta\overline{C}_{\ell 0}$ (ITG-Grace03s-EIGEN-GRACE02S) & $\sigma_{\rm  X}+\sigma_{\rm Y}$ & $f_{\ell}$  (mas yr$^{-1}$)\\
    \hline\hline
    % 2 & $4.81\times 10^{-11}$ & 0\\
    4 & $2.72\times 10^{-11}$ &  $3.9\times 10^{-12}$ & 10.1\\
    6 & $2.35\times 10^{-11}$ &  $2.0\times 10^{-12}$ & 5.1\\
    8 & $1.23\times 10^{-11}$ &  $1.5\times 10^{-12}$ & 0.4\\
    10 & $9.2\times 10^{-12}$ &  $2.1\times 10^{-12}$ & 0.1\\
    12 & $4.1\times 10^{-12}$ &  $1.2\times 10^{-12}$ & -\\
    14 & $5.8\times 10^{-12}$ &  $1.2\times 10^{-12}$ & -\\
    16 & $3.4\times 10^{-12}$ &  $9\times 10^{-13}$ & -\\
    18 & $5\times 10^{-13}$ &  $1.0\times 10^{-12}$ & -\\
    20 & $1.8\times 10^{-12}$ &  $1.1\times 10^{-12}$ & -\\

    \hline
    &    $\delta\mu = 37\%$ (SAV) & $\delta\mu = 24\%$ (RSS) &   \\  %
    \hline
  \end{tabular}

\end{table}
\begin{table}
  \caption{Impact of the mismodelling in the even zonal harmonics estimated in the X=EGM2008 \cite{egm2008} and the Y=EIGEN-GRACE02S \cite{eigengrace02s} models.  The covariance matrix $\sigma$ are calibrated for both EGM2008 and EIGEN-GRACE02S. $\Delta \overline{C}_{\ell 0}$ are larger than the linearly added sigmas for $\ell =4,...,20$, apart from $\ell=18$. }\label{tavola8}
  \begin{tabular}{lccc}
    \hline
    $\ell$ & $\Delta\overline{C}_{\ell 0}$ (EGM2008-EIGEN-GRACE02S) & $\sigma_{\rm  X}+\sigma_{\rm Y}$ & $f_{\ell}$  (mas yr$^{-1}$)\\
    \hline\hline
    % 2 & $4.81\times 10^{-11}$ & 0\\
    4 & $2.71\times 10^{-11}$ &  $8.3\times 10^{-12}$ & 10.0\\
    6 & $2.35\times 10^{-11}$ &  $4.1\times 10^{-12}$ & 5.0\\
    8 & $1.23\times 10^{-11}$ &  $2.7\times 10^{-12}$ & 0.4\\
    10 & $9.2\times 10^{-12}$ &  $2.9\times 10^{-12}$ & 0.1\\
    12 & $4.1\times 10^{-12}$ &  $1.9\times 10^{-12}$ & -\\
    14 & $5.8\times 10^{-12}$ &  $1.8\times 10^{-12}$ & -\\
    16 & $3.4\times 10^{-12}$ &  $1.5\times 10^{-12}$ & -\\
    18 & $5\times 10^{-13}$ &  $1.5\times 10^{-12}$ & -\\
    20 & $1.8\times 10^{-12}$ &  $1.5\times 10^{-12}$ & -\\

    \hline
    &    $\delta\mu = 33\%$ (SAV) & $\delta\mu = 23\%$ (RSS) &   \\  %
    \hline
  \end{tabular}

\end{table}

\begin{table}
  \caption{Bias due to the mismodelling in the even zonals of the models  X = JEM01-RL03B, based on 49 months of GRACE-only data, and Y = AIUB-GRACE01S \cite{aiub}. The latter one was obtained from GPS satellite-to-satellite tracking data and K-band range-rate data out of the
    period January 2003 to December 2003 using the Celestial Mechanics Approach.
    No accelerometer data, no de-aliasing products, and no regularisation was
    applied.
    The $\sigma$ for AIUB-GRACE01S are formal.
    $\Delta \overline{C}_{\ell 0}$ are always larger than the linearly added sigmas.
  }\label{tavolaAIUB1}
  \begin{tabular}{lccc}
    \hline
    $\ell$ & $\Delta\overline{C}_{\ell 0}$ (JEM01-RL03B$-$AIUB-GRACE01S) & $\sigma_{\rm  X}+\sigma_{\rm Y}$ & $f_{\ell}$  (mas yr$^{-1}$)\\
    \hline\hline
    % 2 & $4.81\times 10^{-11}$ & 0\\
    4 & $2.95\times 10^{-11}$ &  $2.1\times 10^{-12}$ & 11\\
    6 & $3.5\times 10^{-12}$  &  $1.3\times 10^{-12}$ & 0.8\\
    8 & $2.14\times 10^{-11}$ &  $5\times 10^{-13}$ & 0.7\\
    10 & $4.8\times 10^{-12}$ &  $5\times 10^{-13}$ & -\\
    12 & $4.2\times 10^{-12}$ &  $5\times 10^{-13}$ & -\\
    14 & $3.6\times 10^{-12}$ &  $5\times 10^{-13}$ & -\\
    16 & $8\times 10^{-13}$ &    $5\times 10^{-13}$ & -\\
    18 & $7\times 10^{-13}$  &    $5\times 10^{-13}$ & -\\
    20 & $1.0\times 10^{-12}$ &    $5\times 10^{-13}$ & -\\

    \hline
    &   $\delta\mu = 26\%$ (SAV)&    $\delta\mu = 23\%$ (RSS)& \\  %
    \hline
  \end{tabular}
\end{table}
\begin{table}
  \caption{Bias due to the mismodelling in the even zonals of the models  X = EIGEN-GRACE02S \cite{eigengrace02s} and Y = AIUB-GRACE01S \cite{aiub}. The $\sigma$ for AIUB-GRACE01S are formal, while those of EIGEN-GRACE02S are calibrated.
    $\Delta \overline{C}_{\ell 0}$ are  larger than the linearly added sigmas for $\ell=4,6,8,16$.
  }\label{tavolaAIUB2}
  \begin{tabular}{lccc}
    \hline
    $\ell$ & $\Delta\overline{C}_{\ell 0}$ (EIGEN-GRACE02S$-$AIUB-GRACE01S) & $\sigma_{\rm  X}+\sigma_{\rm Y}$ & $f_{\ell}$  (mas yr$^{-1}$)\\
    \hline\hline
    % 2 & $4.81\times 10^{-11}$ & 0\\
    4 & $2.98\times 10^{-11}$ &  $6.0\times 10^{-12}$ & 11.1\\
    6 & $2.29\times 10^{-11}$  &  $3.3\times 10^{-12}$ & 5.0\\
    8 & $1.26\times 10^{-11}$ &  $1.9\times 10^{-12}$ & 0.4\\
    10 & $6\times 10^{-13}$ &  $2.5\times 10^{-12}$ & -\\
    12 & $5\times 10^{-13}$ &  $1.6\times 10^{-12}$ & -\\
    14 & $5\times 10^{-13}$ &  $1.6\times 10^{-12}$ & -\\
    16 & $2.9\times 10^{-12}$ &    $1.4\times 10^{-12}$ & -\\
    18 & $6\times 10^{-13}$  &    $1.4\times 10^{-12}$ & -\\
    20 & $2\times 10^{-13}$ &    $1.5\times 10^{-12}$ & -\\

    \hline
    &   $\delta\mu = 34\%$ (SAV)&    $\delta\mu = 25\%$ (RSS)& \\  %
    \hline
  \end{tabular}
\end{table}
The systematic bias evaluated with a more realistic approach is about
3 to 4 times larger than one can obtain by only using this or that
particular model. The scatter is still quite large and differs greatly from that
$5-10\%$ claimed in Ref. \cite{Ciu04}. In particular, it appears that
$J_4$, $J_6$, and to a lesser extent $J_8$, the most
relevant zonals for us owing to their effecta on the combination of
\rfr{combi}, are the most uncertain ones, with discrepancies $\Delta
J_{\ell}$ between different models generally larger than the sum of
their covariances $\sigma_{J_{\ell}}$ whether calibrated or
not.  %This is an important feature because the other alternative combinations proposed involving more satellites \cite{IorAji1,Ves} should be less %affected since they cancel out the impact of $J_4$ and $J_6$ as well.
% Concerning the choice of the models used, it may be argued that it
% is inappropriate to include also solutions not based on GRACE only,
% or not too recent models like, e.g., EIGEN-GRACE01. In fact, there
% are no quantitative reasons for rejecting someone in particular: by
% applying, e.g., the rejection criterion by Chauvenet to the
% various %estimates for $\overline{C}_{40}$ it turns out that it is always $n\gg 0.5$.
% Anyway, table \ref{tavola1}, table \ref{tavola3} and table
% \ref{tavola11} referring to the latest models yield an uncertainty
% $\approx 25\%$ or %more.

Our approach  is valid also for all of the tests performed so far
with the LAGEOS and LAGEOS~II satellites.  Another possible strategy, that takes into account the scatter among the various
solutions, is to compute mean and standard deviation of the
entire set of values of the even zonals for the models considered so
far, degree by degree, and then to take the standard deviations as
representative of the uncertainties $\delta J_{\ell}, \ell =
4,6,8,...$.  This yields $\delta\mu = 15\%$, slightly
larger than that recently obtained in Ref. \cite{Ries08}. But in
evaluating mean and standard deviation for each even zonals,
the authors of Ref. \cite{Ries08} also used global gravity solutions like EIGEN-GL04C and
EIGEN-GL05C which include data from the LAGEOS satellite itself; this
may likely have introduced a sort of favorable \emph{a priori} \virg{imprint}
of the Lense-Thirring effect itself. Moreover, the authors of Ref. \cite{Ries08}
gave only a RSS evaluation of the total bias.

We must also remember to add the further bias due to the
cross-coupling between $J_2$ and the orbit inclination, evaluated to
be about $9\%$ in Ref. \cite{Ior07}.
% \section{Repeatability of the Lense-Thirring test}\lb{repeat}
% The LAGEOS satellites have been accurately tracked for many years by
% the wide network of SLR stations around the world.  Moreover, it can
% be said that, in the slow-motion and weak-field conditions of the
% terrestrial space environment, the measurement of
% the %Lense-Thirring effect can be viewed mainly as a satellite geodesy task, not requiring a highly specialized knowledge in general relativity which may %be lacking in the SLR geodetic community. Thus, analyzing the LAGEOS satellites' data to extract the gravitomagnetic signature should not be, in %principle, a prohibitive task to be implemented. In spite of that, no other tests performed by independent teams, without connections with Ciufolini %and coworkers, following different approaches and using other data analysis techniques have been so far reported in literature. This situation should %be considered unsatisfactory in the sense that if other tests have been so far performed, their outcome should be publicly released, even if, or %rather especially if negative or somehow not conclusive; if, instead, they have not been made at all for some reasons, it would be time to fill such a %gap.
%
\section{A new approach to extract the Lense-Thirring signature from the data}\lb{approach}
The technique adopted so far by the authors of Ref. \cite{Ciu04} and Ref. \cite{Ries08} to
extract the gravitomagentic signal from the LAGEOS and LAGEOS II data
is described in detail in Refs. \cite{LucBal06,Luc07}. The Lense-Thirring
force is not included in the dynamical force models used to fit the
satellites' data. In the data reduction process no dedicated
gravitomagnetic parameter is estimated, contrary to e.g. station
coordinates, state vector, satellites' drag coefficients $C_D$ and
$C_R$, etc.; its effect is retrieved with a sort of post-post-fit
analysis in which the time series of the computed\footnote{The
  expression ``residuals of the nodes'' is used, strictly speaking, in
  an improper sense because the Keplerian orbital elements are not
  directly measured quantities.} ``residuals'' of the nodes with the
difference between the orbital elements of consecutive arcs, combined
with \rfr{combi}, is fitted with a straight line.

In order to enforce the reliability of the ongoing test it would be
desirable to follow other approaches as well. For instance,
the gravitomagnetic force could be modelled in terms of a dedicated
solve-for parameter (not necessarily the usual PPN $\gamma$ one) which
could be estimated in the least-squares sense along with all the other
parameters usually determined, and the resulting correlations among
them could be inspected. Or, one could consider the changes in the
values of the complete set of the estimated parameters with and
without the Lense-Thirring effect.

A first, tentative step towards the implementation of a similar
strategy with the LAGEOS satellites in term of the PPN parameter
$\gamma$ has been recently taken in Ref. \cite{Poz}.
%
% In regard to possible other approaches which could be followed, it
% would be useful to estimate (in the least square sense), among other
% solve-for %parameters, a purely phenomenological correction to the LAGEOS/LAGEOS II node precessions as well and combine them according to \rfr{combi}. %Something similar has been done-although for different scopes-for the perihelia of the inner planets of the Solar System \cite{Pit05} and the %periastron of the pulsars \cite{Kra06}. To be more definite, various solutions with a complete suite of dynamical models, apart from the %Lense-Thirring effect itself, should be produced in which one inserts a solve-for parameter, i.e. a correction (with respect to what modelled) to the %node precessions.  One could see how the outcome varies by changing the data sets and/or the parameters to be solved for.
% Maybe it could be done for each arc, so to have a collection of such
% node extra-rates. Such a strategy would be much more
% model-independent and
% would %be different with respect to the previously suggested way about a Lense-Thirring-dedicated parameter to be estimated along with all the zonals in a %new global solution for the gravity field \cite{Nor01} incorporating the gravitomagnetic component as well; instead, in all the so far produced %global gravity solutions no relativistic parameter(s) have been included in the set of the estimated ones.
% \end{itemize}
\section{Conclusions}
In this paper we have shown how the so far published evaluations of
the total systematic error in the Lense-Thirring measurement with the
combined nodes of the LAGEOS satellites due to the classical node
precessions induced by the even zonal harmonics of the geopotential
are likely optimistic. Indeed, they are all based on the use of elements from the
covariance matrix, more or less reliably calibrated, of
various Earth gravity model solutions used one at a time separately in
such a way that the model X yields an error of $x\%$, the model Y
yields an error $y\%$, etc. Instead, comparing the estimated values of
the even zonals for different pairs of models allows for a much more
realistic evaluation of the real uncertainties in our knowledge of the
static part of the geopotential. As a consequence, the bias in the
Lense-Thirring effect measurement is about three to four times larger than
that so far claimed, amounting to tens of parts per cent (37$\%$ for the
pair EIGEN-GRACE02S and ITG-GRACE03s, about 25--30$\%$ for the other
most recent GRACE-based solutions).
 %
% We have also pointed out that, until now, no other tests of the
% Lense-Thirring effect have been performed by independent teams,
% although it would
% be, %at least in principle, a relatively not too demanding task in view of the wide dissemination of SLR stations, for which the LAGEOS satellites are %important targets since long time, and of the freely available softwares to perform the data reduction process.
 %

Finally, we have pointed out the need of following different
strategies in extracting the Lense-Thirring pattern from the data; for
instance by explicitly modelling it in fitting the SLR data of LAGEOS
and LAGEOS II, and estimating the associated solve-for parameter in a
least-square sense along with the other parameters usually determined.
% \section*{Acknowledgments}
% I am grateful to J Ries (CSR) and M Watkins (JPL) for having
% provided me with the spherical harmonics' coefficients of the GGM03S
% and JEM01-RL03B %models.
% -----------------------------------------

\end{document}